\documentstyle[12pt,aaspp4,psfig,natbib]{article}

\newcommand{\vy}[2]{#1_{\scriptscriptstyle #2}}
\newcommand{\Ly}{Ly$\alpha$}

\def\gtorder{\mathrel{\raise.3ex\hbox{$>$}\mkern-14mu
             \lower0.6ex\hbox{$\sim$}}}
\def\ltorder{\mathrel{\raise.3ex\hbox{$<$}\mkern-14mu
             \lower0.6ex\hbox{$\sim$}}}
\def\proptwid{\mathrel{\raise.3ex\hbox{$\propto$}\mkern-14mu
             \lower0.6ex\hbox{$\sim$}}}
\def\0946{PG~0946+301}

\def\arcsec{\ifmmode '' \else $''$\fi}

\def\arcsecpoint{\ifmmode ''\!. \else $''\!.$\fi}

\def\kms{\ifmmode {\rm km\ s}^{-1} \else km s$^{-1}$\fi}
\def\Msun{\ifmmode {\rm M}_{\odot} \else M$_{\odot}$\fi}
\def\Lsun{\ifmmode {\rm L}_{\odot} \else L$_{\odot}$\fi}
\def\Zsun{\ifmmode {\rm Z}_{\odot} \else Z$_{\odot}$\fi}

\def\ergscm2{ergs\,s$^{-1}$\,cm$^{-2}$}
\def\icm3{{\rm cm}^{-3}}
\def\icm2{{\rm cm}^{-2}}
\def\qo{\ifmmode q_{\rm o} \else $q_{\rm o}$\fi}
\def\Ho{\ifmmode H_{\rm o} \else $H_{\rm o}$\fi}
\def\ho{\ifmmode h_{\rm o} \else $h_{\rm o}$\fi}

\def\vFWHM{\ifmmode v_{\mbox{\tiny FWHM}} \else
            $v_{\mbox{\tiny FWHM}}$\fi}
\def\CCF{\ifmmode F_{\it CCF} \else $F_{\it CCF}$\fi}
\def\ACF{\ifmmode F_{\it ACF} \else $F_{\it ACF}$\fi}
\def\Halpha{\ifmmode {\rm H}\alpha \else H$\alpha$\fi}
\def\Hbeta{\ifmmode {\rm H}\beta \else H$\beta$\fi}
\def\Hgamma{\ifmmode {\rm H}\gamma \else H$\gamma$\fi}
\def\Hdelta{\ifmmode {\rm H}\delta \else H$\delta$\fi}
\def\Lya{\ifmmode {\rm Ly}\alpha \else Ly$\alpha$\fi}
\def\Lyb{\ifmmode {\rm Ly}\beta \else Ly$\beta$\fi}
\def\Lyg{\ifmmode {\rm Ly}\beta \else Ly$\gamma$\fi}
\def\hi{H\,{\sc i}}

\def\heii{He\,{\sc ii}}

\def\cii{C\,{\sc ii}}
\def\ciii{\ifmmode {\rm C}\,{\sc iii} \else C\,{\sc iii}\fi}
\def\civ{\ifmmode {\rm C}\,{\sc iv} \else C\,{\sc iv}\fi}

\def\cvi{C\,{\sc vi}}

\def\nv{N\,{\sc v}}
\def\nvi{N\,{\sc vi}}

\def\o5007{[O\,{\sc iii}]\,$\lambda5007$}

\def\ovi{O\,{\sc vi}}

\def\siiv{Si\,{\sc iv}}

\def\siIII{Si\,{\sc iii}}

\def\siv{S\,{\sc iv}}

\def\svi{S\,{\sc vi}}

\def\o{\o}
\begin{document}

\title{CHEMICAL ABUNDANCES IN AGN ENVIRONMENT: \\
       X-RAY/UV CAMPAIGN ON THE MRK 279 OUTFLOW
}


\author{Nahum Arav\altaffilmark{1}, Jack R. Gabel\altaffilmark{2},
Kirk T. Korista\altaffilmark{3}, Jelle S. Kaastra\altaffilmark{4},
Gerard A. Kriss\altaffilmark{5,6}, Ehud Behar\altaffilmark{7}, Elisa
Costantini\altaffilmark{4,8}, C.~Martin Gaskell\altaffilmark{9}, Ari
Laor\altaffilmark{7}, Nalaka Kodituwakku\altaffilmark{3}, Daniel
Proga\altaffilmark{10}, Masao Sako\altaffilmark{11}, Jennifer
E. Scott\altaffilmark{12}, Katrien C. Steenbrugge \altaffilmark{13}}

\begin{abstract}

We present the first reliable determination of chemical abundances in
an AGN outflow. The abundances are extracted from the deep and
simultaneous FUSE and HST/STIS observations of Mrk~279.  This data set
is exceptional for its high signal-to-noise, unblended doublet troughs
and little Galactic absorption contamination.  These attributes allow
us to solve for the velocity-dependent covering fraction, and
therefore obtain reliable column densities for many ionic species. For
the first time we have enough such column densities to simultaneously
determine the ionization equilibrium and abundances in the flow.  Our
analysis uses the full spectral information embedded in these
high-resolution data. Slicing a given trough into many independent
outflow elements yields the extra constraints needed for a physically
meaningful abundances determination.  We find that relative to solar
the abundances in the Mrk~279 outflow are (linear scaling): carbon
2.2$\pm$0.7, nitrogen 3.5$\pm$1.1 and oxygen 1.6$\pm$0.8.  Our
UV-based photoionization and abundances results are in good agreement
with the independent analysis of the simultaneous Mrk~279 X-ray
spectra.  This is the best agreement between the UV and X-ray analyses
of the same outflow to date.

\end{abstract}


\altaffiltext{0}{Based on observations made with the NASA/ESA {\it Hubble 
Space Telescope} and the NASA-CNES-CSA {\it Far Ultraviolet Spectroscopic Explorer},
and obtained at the Space Telescope Science Institute, which is operated by the 
Association of Universities for Research in Astronomy, Inc. under NASA 
contract NAS~5-26555.}

\altaffiltext{1}{Center for Astrophysics and Space Astronomy, 
University of Colorado, 389 UCB, Boulder CO 80309-0389; 
arav@colorado.edu}
\altaffiltext{2}{Department of Physics, Creighton University, 
2500 California Plaza, Omaha NE 68178;
JackGabel@creighton.edu}
\altaffiltext{3}{Department of Physics, Western Michigan University, 
Kalamazoo, MI 49008; korista@wmich.edu}
\altaffiltext{4}{SRON National Institute for Space Research, Sorbonnelaan 2,
3584 CA Utrecht, The Netherlands; J.S.Kaastra@sron.nl, e.costantini@sron.nl, 
K.C.Steenbrugge@sron.nl}
\altaffiltext{5}{Space Telescope Science Institute, 3700 San Martin Drive, 
Baltimore, MD 21218; gak@stsci.edu}
\altaffiltext{6}{Center for Astrophysical Sciences, Department of Physics and 
Astronomy, The Johns Hopkins University, Baltimore, MD 21218}
\altaffiltext{7}{Department of Physics, Technion, Haifa 32000, Israel;
behar@physics.technion.ac.il, laor@physics.technion.ac.il}
\altaffiltext{8}{Astronomical Institute, University of Utrecht, PO Box 80 000,
3508 TA Utrecht, The Netherlands}
\altaffiltext{9}{Department of Physics and Astronomy, University of Nebraska,
Lincoln NE 68588-0111; mgaskell1@unl.edu}
\altaffiltext{10}{Physics Department, University of Nevada, Las Vegas, 
 Box 454002 Las Vegas, NV 89154-4002 ; dproga@physics.unlv.edu}
\altaffiltext{11}{Dept. of Physics and Astronomy,
209 South 33rd Street, Philadelphia, PA 19104; masao@sas.upenn.edu,}
\altaffiltext{12}{Dept. of Physics, Astronomy, and Geosciences
Towson University; jescott@towson.edu}
\altaffiltext{13}{Department of Physics, University of
Oxford, Keble Road, Oxford, OX1 3RH, UK; katrien@head.cfa.harvard.edu}

\clearpage

\section{INTRODUCTION}

Active galactic Nuclei (AGN) provide a vital probe to the early
history of chemical enrichment in the universe (Shields 1976; Hamann
\& Ferland 1992; ,Ferland et~al.\ 1996; Dietrich et~al.\ 2003).  Their
brightness allow us to see chemically processed environments when the
universe was less than 7\% of it's current age (Fan et~al.\ 2004),
thus giving us insights into early star formation and galaxy
evolution.  The spectra of the highest redshift AGN are very similar
to those of AGN in the local universe (Fan et~al.\ 2004), suggesting
that chemical enrichment of their environments operates on short
cosmological time-scales.  This feature makes local AGN a good probe
of chemical processing in the early universe.

There are two main tracks for determining abundances in AGN: Broad
emission lines (BELs) are seen in most AGN and are known to be formed
in close proximity of the nucleus (0.01--0.1 pc.; Kaspi et~al.\ 2005,
and references therein).  Considerable effort was put into trying to
determine abundances in AGN by studying the relative strength of BELs
in individual as well as composite spectra (e.g., Shields 1976;
Baldwin \& Netzer 1978; Hamann \& Ferland 1992; Ferland et~al.\ 1996;
Dietrich et~al.\ 1999, 2003).  These studies suggest that the
metalicity of the BEL region (BELR) is super solar.  However, BEL
abundances studies are only relative in the sense that they do not
measure the ratio of heavy elements to hydrogen directly.  Ratios of
nitrogen to carbon and oxygen BELs are measured and then converted to
metalicity (Z) using the expected secondary neucleosynthesis of
nitrogen, which predicts N/O and N/C~$\propto Z$ or N/H~$\propto Z^2$
(Hamann \& Ferland 1999).  This assumption weakens the robustness of
the BEL metalicity claims in principle. In the best case where \hi\
measurements of galactic environment are available in a high redshift
galaxy, this assumption is not verified (Pettini et~al.\ 2002).
Furthermore, there are considerable systematic issues that affect
these studies.  The BELR is spatially stratified and likely to have a
wide range of densities and temperatures (Nagao et~al.\
2006). Therefore, different BELs can arise from largely different
regions, which complicates the use of BEL line flux ratios as
abundance indicators.  Radiation transfer inside the opaque BEL
material is a difficult problem that further complicates interpreting
observed BEL ratios (Netzer 1990).

The second track for determining abundances is using absorption lines
associated with AGN outflows.  In principal, absorption line studies
allow for absolute abundances measurements since the hydrogen Lyman
series troughs can yield direct ratios of hydrogen to heavy elements.
AGN outflows are evident by resonance
line absorption troughs, which are blueshifted with respect to the
systemic redshift of their emission counterparts. In Seyfert galaxies
velocities of several hundred \kms\ (Crenshaw et~al.\ 1999; Kriss
et~al.\ 2000) are typically observed in both UV resonance lines (e.g.,
\civ~$\lambda\lambda$1548.20,1550.77,
\nv~$\lambda\lambda$1238.82,1242.80,
\ovi~$\lambda\lambda$1031.93,1037.62 and \Ly), as well as in X-ray
resonance lines (Kaastra et~al.\ 2000, 2002; Kaspi et~al.\ 2000,
2002). Similar outflows (often with significantly higher velocities)
are seen in quasars which are the luminous relatives of Seyfert
galaxies (Weymann et~al.\ 1991; Korista, Voit, Morris, \& Weymann
1993; Arav et~al.\ 2001a). Distances of the outflows from the central
source can range from smaller than BELR distances (QSO 1603; Arav et
al 2001); tens of pc. (NGC 3783; Gabel et~al.\ 2005b); 1000 pc. (QSO
1044; de Kool et~al.\ 2001).  Thus, in most cases the outflows are
associated with material in the vicinity of the AGN, and can be used
as diagnostics for the chemically enriched environment at the center
of galaxies.

A main obstacle in determining abundances using outflow absorption
lines is obtaining reliable measurements of the absorption column
densities from the troughs.  In the last few years our group (Arav
1997; Arav et~al.\ 1999a; Arav et~al.\ 1999b; de~Kool et~al.\ 2001;
Arav et~al.\ 2002, 2003) and others (Barlow 1997, Telfer et~al.\ 1998,
Churchill et~al.\ 1999, Ganguly et~al.\ 1999) have shown that in
quasar outflows most lines are saturated even when not black.  As a
consequence the apparent optical depth method, which stipulates that
the optical depth $\tau_{ap}\equiv-\ln(I)$, where $I$ is the residual
intensity in the trough, is not a good approximation for outflow
troughs. In addition, using the doublet method (Barlow 1997, Hamann et
al.1997) we have shown that in many cases the shapes of the troughs
are almost entirely due to changes in the line of sight covering as a
function of velocity, rather than to differences in optical depth
(Arav et~al.\ 1999b; de~Kool et~al.\ 2001; Arav et~al.\ 2001a; Gabel
et~al.\ 2005a). Gabel et~al.\ (2003) show the same effect in the
outflow troughs of NGC~3783, as does Scott et~al.\ (2004) for Mrk~279.
As a consequence, the column densities inferred from the depths of the
troughs are only lower limits.

In order to measure reliable column densities we found it necessary to
use a two step combination: First, we developed analysis methods that
can disentangle the covering factors from the optical depth (Gabel et
al.\ 2003, 2005a). Second, these methods depend on having
high-resolution ($\gtorder$ 20,000) and high signal-to-noise (S/N
$\gtorder$ 20) spectral data of outflow troughs.  Furthermore, we
critically rely on fully-resolved uncontaminated doublet and multiple
troughs.  Mrk~279 is the optimal target for such analysis, since it
allows us to obtain data with the above specification on the \civ,
\nv, and \ovi\ doublets, as well as on several Lyman series troughs.
The latter are crucial for any abundances determination (see
\S~5.1). Before applying this analysis to the Mrk~279 data, we tested
the validity of our absorption model.  The main alternatives are
inhomogeneous absorber models (de Kool, Korista \& Arav 2002).  We
tested the main variants of inhomogeneous absorber models on the
Mrk~279 data set and concluded that the outflow cannot be fitted well
with these models (Arav et~al.\ 2005; see \S~5.2.1).  Following these
tests we used the partial covering absorber model to extract the
column densities of all the observed ions in the Mrk~279 outflow (Gabel
et~al.\ 2005a).

In this paper we present the determination of chemical abundances in
the AGN outflow emanating from Mrk~279, using the high-quality
simultaneous UV data sets from HST/STIS and FUSE.  Our procedure for
determining the abundances goes as follows: We treat each spectral
resolution element as an independent measurement (see \S~5.1 for
discussion).  After determining the spectral energy distribution (SED)
appropriate for Mrk~279 (\S~2.2), we solve for the ionization
parameter ($U$) and the total hydrogen column density ($\vy{N}{H}$),
independently for each of our 15 resolution elements (\S~2.3).  This
is done by comparing the measured column densities with
photoionization grid models, assuming a given set of CNO abundances.
Each resolution element is allowed to have a different combination of
$U$ and $\vy{N}{H}$.  However, we require the same set of CNO
abundances for all resolution elements.  The CNO composition of the
outflow is determined to be the abundance set that yields the best fit
for all 15 resolution elements (\S~3).  Section 4 is devoted to a
comparison of the photoionization and abundances results presented
here with the independent analysis of the simultaneous Mrk~279 X-ray
spectra. In \S~5 we discuss our results.
The appendix gives the details for the extraction of ionic column
densities from the observed troughs, with emphasis on the associated
errors.


\section{PHOTOIONIZATION ANALYSIS}

\subsection{The Mrk 279 Data Set}

On May 2003 we obtained simultaneous X-ray and UV observations of the
Mrk~279 AGN outflow. (Description of the UV observations is found in
Gabel et~al.\ 2005a, and the X-ray observations in Costantini et~al.\
2006.)  The 92 ks FUSE data have a spectral resolution of 20,000 and
cover the observed wavelength 905--1187\AA. These data yielded the
highest quality \ovi\ trough-spectrum of any AGN outflow to date. The
16 orbits HST/STIS/E140M observation have a spectral resolution of
40,000 and cover the observed wavelength 1150--1730\AA. These data
yielded high quality \civ\ and \nv\ troughs. The combined HST/FUSE
spectrum gives high signal-to-noise \Lya, \Lyb\ and \Lyg\ troughs.
Compared with other well-studied AGN outflow targets, the Galactic \hi\
column density in the direction of Mrk~279 is quite low 1.6 compared
with 4.4 and 8.7 for Mrk~509 and NGC~3783, respectively (in units of
$10^{20}$ cm$^{-2}$; Hwang \& Bowyer 1997; Kaspi et al 2002).
Therefore, the outflow troughs of Mrk~279 are significantly less
contaminated with galactic absorption features than the other noted
targets.

In Gabel et~al.\ (2005a) we used a global fitting technique to extract
three velocity-dependent quantities from the data: Continuum covering
factor, BELR covering factor and column densities $N_{ion}$ for all
the detected ions.  An important finding was that the outflow fully
covers the continuum both for the Lyman series troughs and for the CNO
doublets.  In this paper we refine the $N_{ion}$ measurements by using
this finding.  Once the continuum coverage is known to be 100\%, the
doublet equations are reduced to two unknowns (BELR covering factor
and optical depth), which we can solve for independently for each ion,
since we have two residual intensity equations.  These solutions are
more physical and their errors are easier to estimate.  In the
appendix we give the details of the new solution while for the rest of
the paper we use this improved set of $N_{ion}$ determinations.  The
difference between the Gabel et al (2005a) and the current $N_{ion}$
measurements can be seen in the first figure of the appendix (Fig.\
\ref{fig:N_ion_errors}). We point out that the main difference is in
the size of the error bars, where we put much effort into obtaining
physically accurate errors since the analysis is crucially dependent
on these.
 
\subsection{Photoionization Modeling and SED}

We use simple slab photoionization models (using the code  Cloudy;
Ferland 1998).  Such models are commonly used in the study of quasar
outflows (e.g., Weymann, Turnshek, \& Christiansen 1985; Arav,  Li \& Begelman
1994; Hamann 1996; Crenshaw \& Kraemer 1999; Arav et~al.\ 2001b;
Gabel et~al.\ 2006) and assume that the absorber consists of a constant
density slab irradiated by an ionizing continuum.  The two main
parameters in these models are the thickness of the slab as measured
by the total hydrogen column density ($\vy{N}{H}$) and the ionization
parameter $U$ (defined as the ratio of number densities between
hydrogen ionizing photons and hydrogen in all forms).  

\begin{figure}[h]
\centerline{\psfig{file=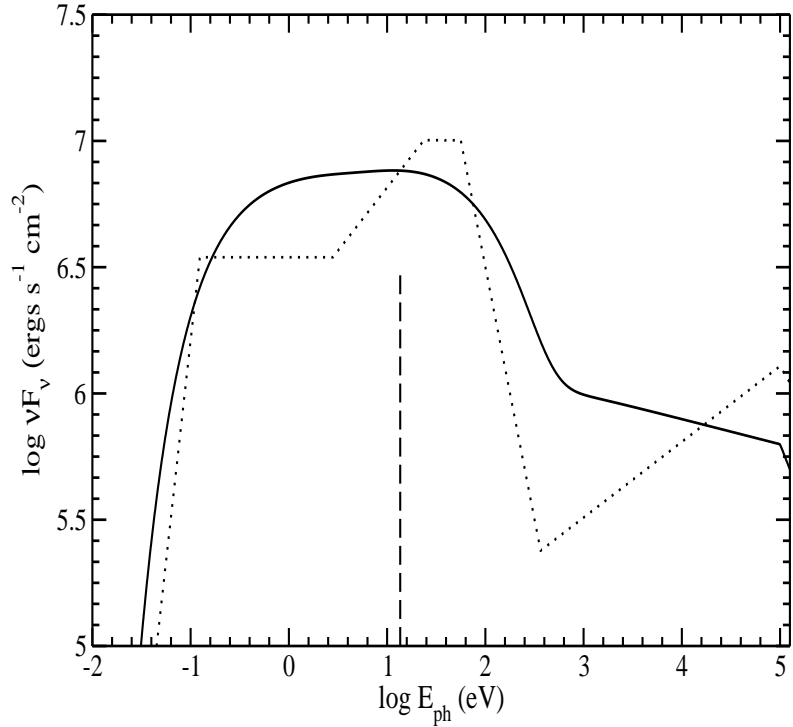,angle=-90,height=12.0cm,width=12.0cm}}
\caption{SED used for our analysis (solid line) compared with the
canonical Mathews-Ferland spectrum. The two SEDs are normalized to the 
same flux at 1 Rydberg (shown by the dashed line, 1 Rydberg is equivalent
to 13.6 eV, or 912 \AA).
The Mrk~279 model incident  SED was constructed to smoothly 
connect the FUSE spectrum near rest frame 900A to the Chandra LETGS 
spectrum near 200 eV. (see text for details). 
Compared with the Mathews-Ferland SED, the Mrk~279 SED is somewhat harder.
}
\label{fig:SED}
\end{figure}

The spectral energy distribution (SED) of the incident flux is also
important in determining the ionization equilibrium. The ionizing SED
of Mrk~279 is arguably the most tightly constrained of any AGN to
date.  FUSE observations set the flux at $\sim900$\AA\ in the object's
rest frame and the LETGS on board Chandra {\em simultaneously}
provided reliable X-ray coverage for energies greater than 200
eV. Even so, some ambiguity remains due to the gap between 13.8 eV and
200 eV (see fig \ref{fig:SED}). We opted to continue the slope of the
FUSE continuum and gradually steepen it until it matches the soft
X-ray continuum.  We chose this shape rather than connecting these two
points with a simple power-law, since we consider it unlikely that a
spectral break falls immediately at the edge of the observed UV data.
We note that the 13.8-200 eV continuum would be somewhat softer with a
simple power-law connecting the UV and X-ray data, and elaborate on
the ramifications of such an SED in the Discussion.

To generate the Mrk~279 SED shown in fig \ref{fig:SED}, we used the Cloudy
`agn' command that generates the sum of two continua,
 $F_{\nu} = A\nu^{\alpha_{uv}}exp(-h\nu/kT_{uv}) + B\nu^{\alpha_x}$. 
The first of these represents the optical/UV
bump, and this command also imposes a low energy exponential cutoff at
1 micron. We chose $\alpha_{uv} = -1.0$ and $kT_{uv} = 136$ eV to
smoothly and conservatively span the 900~\AA\/ $-$ 200~eV spectral
region. The second of the continua represents a high energy (X-ray)
power law that lies between 1 Ryd and 100 keV with cutoffs at both
ends. We chose a slope to match the observed one, $\alpha_{x}= -1.1$,
and normalized this component's strength relative to the UV bump with
an $\alpha_{ox} = -1.35$. The Cloudy input command string that
generated this continuum is `agn 6.2 -1.35 -1.0 -1.1', where the first
parameter is $\log T_{uv}$, followed by $\alpha_{ox}$, $\alpha_{uv}$,
and $\alpha_{x}$.


\subsection{Velocity Dependent Grid Models}

We base our photoionization analysis, and ultimately the abundance
determinations on the grid-of-models approach created by Arav et~al.\
(2001b).  For a given $N_{ion}$ we ran a grid of models spanning three
orders of magnitude in both $\vy{N}{H}$ and $U$.  For these models we
use the Mrk~279 SED described above and assume solar abundances, as
given by the CLOUDY code, where the carbon and oxygen abundances are
taken from Allende Prieto et al. (2002, 2001), and the nitrogen
abundance is taken from Holweger (2001).  The solar abundance of these
elements compared to hydrogen are: C=$2.45\times10^{-4}$,
N=$8.5\times10^{-5}$, O=$4.9\times10^{-4}$.  We then determine which
combinations of $\vy{N}{H}$ and $U$ reproduce the observed $N_{ion}$
and then plot the ion-curve, which is the locus of these points on the
$\vy{N}{H}$/$U$ plane.  We do that for each measured $N_{ion}$.
Ideally, a single value of $\vy{N}{H}$ and $U$ should be able to
reproduce all the $N_{ion}$ constraints.  This is based on the
assumption that the measured $N_{ion}$ arise from a parcel of gas with
uniform density that is irradiated by the given SED.
On the $\vy{N}{H}$/$U$ plane this single value solution will be
represented by the crossing of all the ion-curves at a single point.

A new feature that adds considerable diagnostic power to these models,
is their velocity dependence.  Since the $N_{ion}$ measurements are for
a specific velocity, we obtain a grid model for each element of
resolution along the trough.  There is no a-priori reason why material
flowing in different velocities should have the same ionization
structure or have the same column density per unit velocity (see \S~5.1).
Separating the absorption trough into individual resolution elements
therefore allows for a more accurate solution of the ionization
structure of the flow, and as we show below, is the key for precise
abundances determination.

\begin{figure}[ht]
\centerline{\psfig{file=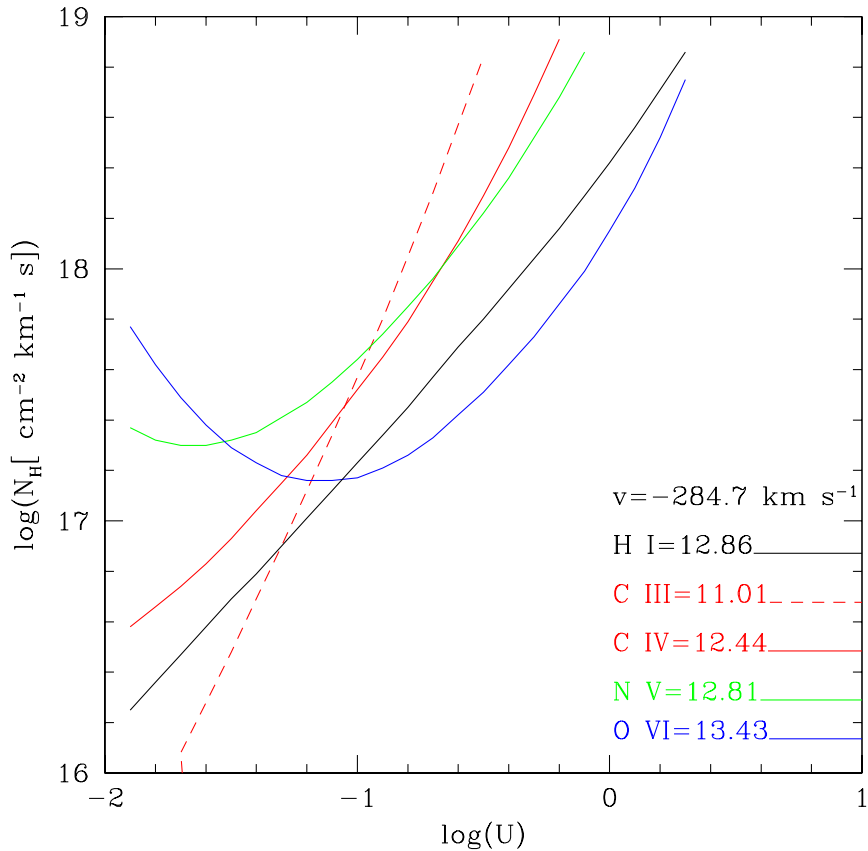,angle=0,height=12.0cm,width=12.0cm}}
\caption{Curves of constant ionic column density plotted on the plane
of total hydrogen column density ($\vy{N}{H}$) per unit velocity of
the slab vs. the ionization parameter of the incident radiation ($U$),
using solar abundances and an SED tailored for these specific
observations of Mrk~279 (see Fig.\ \ref{fig:SED}).  The inserted Table shows 
measured  $\log(N_{ion})$ [per unit velocity] for the resolution-element
centered around the $-285$ \kms\ outflow velocity. }
\label{fig:nh_u_285_c_nion}
\end{figure}

Figure \ref{fig:nh_u_285_c_nion} shows the grid model for the 
resolution-element centered around the $-285$ \kms\ outflow velocity.
All the measured ions are represented. The upper limits from \siiv,
\siv, \svi\ and \cii, do not add meaningful constraints and are
consistent with the results of our analysis.  An important advantage over the
analysis of the PG~0946+301 spectra (Arav et~al.\ 2001), is that all the 
$N_{ion}$ shown on figure \ref{fig:nh_u_285_c_nion} are actual measurements
and not lower limits, as was the case for most ions in PG~0946+301.
 
\subsection{Measurement Errors and $\chi^2$ Analysis}

In measuring the $N_{ion}$ for the outflow troughs in Mrk~279 we
accounted for most of the systematic issues that plagued previous
measurements. Unblended doublet and multiplet troughs were used, 
without which reliable outflow $N_{ion}$ cannot be measured in principle.
Both the continuum and BELR covering factors were taken
into consideration, and $N_{ion}$ as a function of velocity was extracted.
We therefore possess the first comprehensive set of reliable $N_{ion}$
measurements for an AGN outflow.     

Extracting meaningful constraints on the ionization equilibrium and
abundances from the measured $N_{ion}$ requires having a reliable data
base of the associated errors.  Estimating physical errors for the
$N_{ion}$ measurements is a complicated process that is fully
described in the Appendix.  Here we give a brief description of this
process and a qualitative explanation for the different error values
of the individual ions. For all ions except \ciii\ the fitting
procedure simultaneously finds the optimal covering factor and optical
depth at a given velocity point.  Once the optimal combination is
found, we derive the formal error of the parameters (see Appendix for
full discussion).  Our approach is verified by the close match between
the resultant flux deviations and the empirical flux errors in both
positive and negative directions (see the second figure in the
Appendix, Fig~7).

The errors are not symmetrical.  Typically, the $+$ error is larger
than the $-$ error and often much larger.  This is due to the
exponential dependence of the absorption on the optical depth.  For
example if $\tau=4$ we already absorb 98\% of the flux covered by the
outflow.  Increasing $\tau$ by 2 only absorbs an additional 2\%. A
similar increase in flux requires a decrease of only 0.6 in $\tau$.

\begin{figure}[ht]
\centerline{\psfig{file=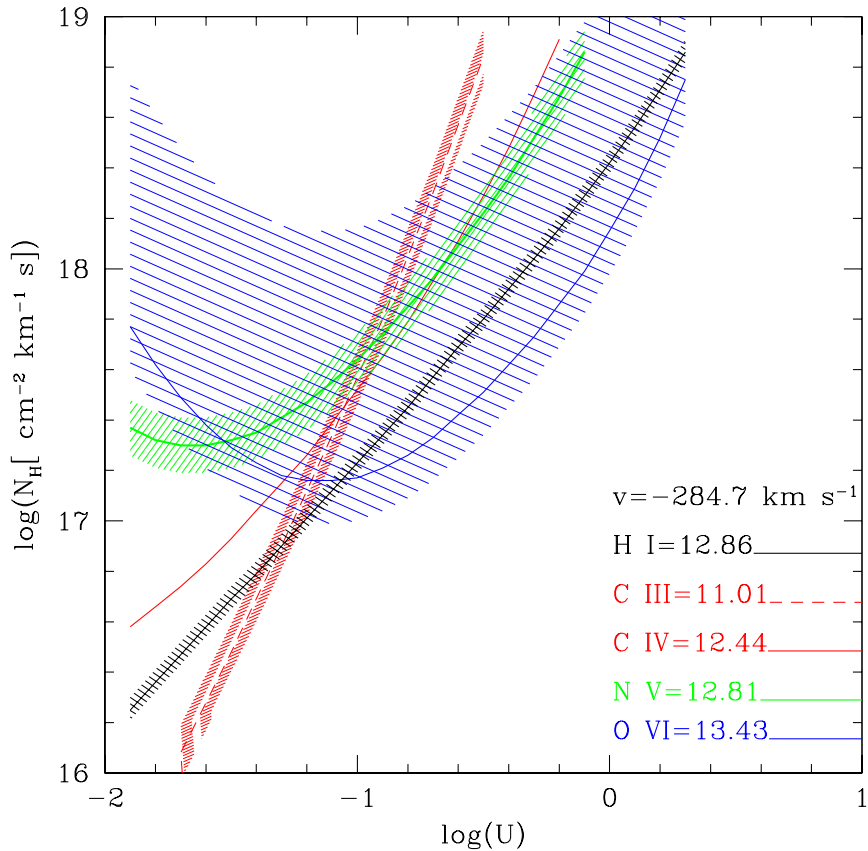,angle=0,height=12.0cm,width=12.0cm}}
\caption{Same as \ref{fig:nh_u_285_c_nion}, but including 1$\sigma$
uncertainties, shown as shaded areas having the same color as their
associated ion curve.   For the sake of clarity, we omitted
the 1$\sigma$ presentation for \civ.  As a figure of merit, in this
velocity plot the 1$\sigma$ interval for \civ\ touches the \hi\ curve
around $\log(U)=-1.4$.}
\label{fig:nh_u_285_e_errors}
\end{figure}

In figure \ref{fig:nh_u_285_e_errors} we show the errors associated
with the photoionization model presented in figure \ref{fig:nh_u_285_c_nion}.
\ovi\ has the largest errors  because at that velocity its troughs are nearly
saturated, and for that reason the asymmetry in its errors is also most
pronounced.  \hi\ exhibits the smallest errors, and therefore supplies
the strongest constraint on the physical solutions for the outflowing
gas.  The \hi\ errors are the smallest because the observations cover
five Lyman series lines and the intrinsic optical depth ratio of the
first three spans a factor of 18.  This, coupled with the fact that the \Lyg\
trough is not close to saturation, yield tight constraints on the
column density.  In contrast the intrinsic optical depth ratio between 
the members of each CNO doublet is 2, yielding significantly
less stringent constraints on their $N_{ion}$.  \ciii\ is a singlet 
and therefore we had to make an assumption regarding its covering fraction.
Once we assumed the \ciii\ coverage to be the same as \civ\  we derived tight
constraints on its $N_{ion}$ since the trough was far from saturation.
Finally, a contributing factor for the larger errors of the CNO doublets  
is that flux errors are associated with both doublet components
while often the difference between the two residual intensities
is small compared to the flux values themselves.   

\begin{figure}[ht]
\centerline{\psfig{file=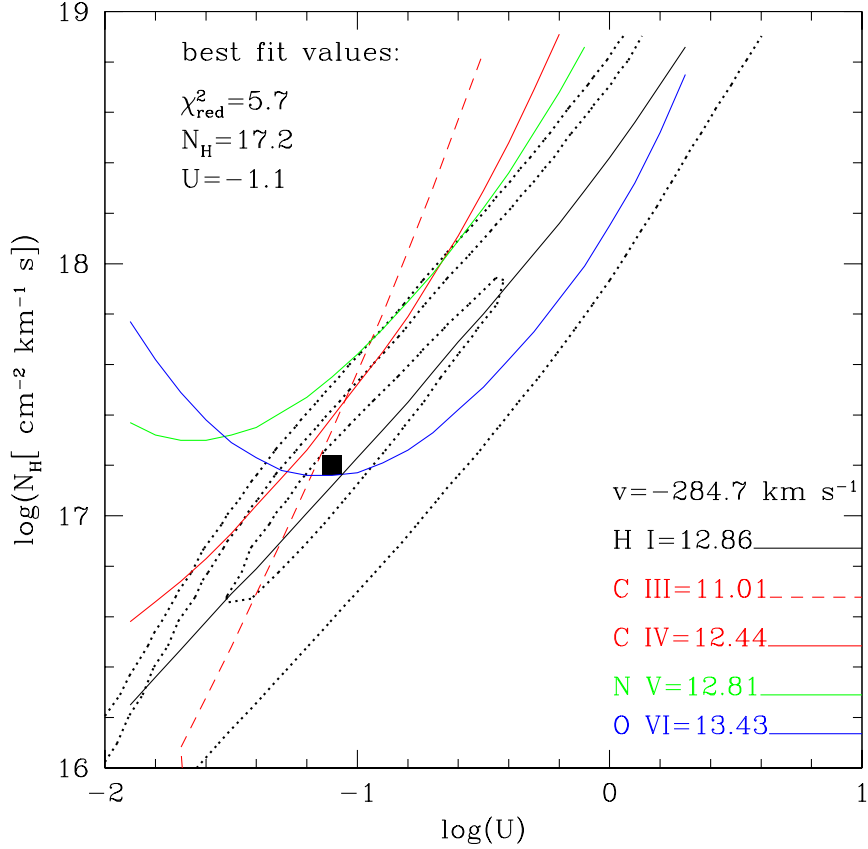,angle=0,height=12.0cm,width=12.0cm}}
\caption{ Similar to Fig.\ \ref{fig:nh_u_285_c_nion} over-plotted with
reduced $\chi^2$ contours (at 12.5,50,112) for an $\vy{N}{H},U$ solution
(dotted lines).  Position of the formal solution is marked by the
filled square and the best fit values are shown at the top left
corner.  From the large value of the reduced $\chi^2$ it is evident
that there is no acceptable $\vy{N}{H},U$ solution for this solar
abundances model.  }
\label{fig:nh_u_285_f_chi2}
\end{figure}

We are now in position to ask the first important physical question:
Do the ion curves in figure \ref{fig:nh_u_285_e_errors} allow for an
acceptable solution for the ionization equilibrium for this particular
velocity slice of the outflow?  That is, is there a combination of
$\vy{N}{H}$ and $U$ that will satisfy all the ion-curve constraints?
It is evident from figure \ref{fig:nh_u_285_e_errors} that this is not
the case, as the error stripes for \hi\ and \nv\ do not overlap on
this grid-model plot.  A physical solution cannot be obtained for this
velocity-slice of the outflow assuming solar abundances.  In figure
\ref{fig:nh_u_285_f_chi2} we show the formal $\chi^2$ solution for the
best fit values of $\vy{N}{H}$ and $U$ of this velocity slice.  The
reduced $\chi^2=5.7$ formally shows that the model does not
yield a statistically valid fit for the data.  
of the 15 slices, 13 yield  reduced $\chi^2$ between 2-10, and only two 
have reduced $\chi^2$ smaller than 2.

\section{ABUNDANCES DETERMINATION}

\subsection{Method}

The most plausible way to obtain a physical solution for $\vy{N}{H}$
and $U$ in figure \ref{fig:nh_u_285_e_errors} is to drop the
assumption of solar abundances.  On our parameter space plots an
increase in abundance is equivalent to lowering the entire related
ion-curve, since less $\vy{N}{H}$ is needed for the same amount of the
said element.  We can see from figure \ref{fig:nh_u_285_e_errors} that
if the nitrogen abundance is 3 times higher than solar, the \nv\
ion-curve will be lower by the same amount and the error stripes for
\hi\ and \nv\ will comfortably overlap in parts of the parameter
space.  

For one velocity slice we can always achieve a perfect solution (exact
crossing of the ion curves) by allowing the CNO abundances relative to
hydrogen to be free parameters.  Such a solution might be suggestive,
but it is considered weak since we use three free parameters to make
the 5 ion-curves cross at the same point.  As can be seen in figure
\ref{fig:nh_u_285_c_nion}, this is always the case.  As long as we
specify the value of $U$ to be that of the \ciii\ and \civ\ ion-curves
crossing point, the three free abundances parameters allow us to bring all
CNO ion-curves to meet at that $U$ value on top of the \hi\ curve.

The way to obtain a robust rather than suggestive abundance
determinations for the outflow is to use the velocity information.  So
far we have only dealt with a single velocity-slice.  It is reasonable
to expect that each velocity-slice will have an independent ionization
solution, i.e., its own $\vy{N}{H},U$ values.  The number density and
column density of the outflow at each resolution element are
determined by the flow's dynamics, which can be quite complicated
(Proga 2005).  However, it is likely that an outflow component would
have the same abundances at all velocities.  Therefore, the
assumptions we make are: 1) each velocity-slice has an independent
$\vy{N}{H},U$ ionization solution; 2) all velocity slices of the same
outflow component have the same chemical abundances.

\subsection{Velocity Coverage}

In Mrk~279 the outflow spans the velocity range $-200$ to $-540$ \kms.
As described in Gabel et~al.\ (2005a) the velocity range $-300$ to
$-540$ \kms\ is contaminated by unrelated absorption (probably from
the companion galaxy of Mrk 279, see Scott et~al.\ 2004), which is
detected in lines from low-ionization species (\cii, \ciii, \siIII),
and especially in the \hi\ Lyman series.  For this reason we cannot
use the \hi\ trough measurements in the range $-300$ to $-540$ \kms\
in our ionization and abundance analysis of the flow.  We therefore
concentrate on the $-200$ to $-300$ \kms\ range where we have high
quality uncontaminated measurements for \hi, \ciii, \civ, \nv\ and
\ovi.  The HST STIS 140M grating yields 15 independent resolution
elements for this range.  We use the \nv\ trough spectrum to fix the
velocity points of these elements and interpolate all the other
trough's spectra to that velocity scale.  FUSE's resolution
$(\lambda/\Delta\lambda)$ is only half that of the STIS E140M grating,
and we interpolated them on the same velocity scale.  We used the STIS
rather than the FUSE resolution for the following reasons: a)
Maintaining all the physical information of the STIS troughs.  b) The
\ovi\ doublet covered by FUSE was the most saturated and yielded the
least constraining measurements, therefore an oversampling by a factor
of two will not affect the physical conclusion significantly.  c) The
\hi\ measurements arise from one STIS line (\Lya) and two FUSE lines
(\Lyb\ and \Lyg) and the derived $N_{\hi}$ is smooth as a function of
velocity (See Fig.\ \ref{fig:N_ion_errors}).  Thus, we preserve the
STIS velocity resolution for \Lya, while the smoothness of the
measurements minimizes the effects of oversampling the FUSE data on
the physical conclusions, since no structure is seen on small velocity
scales.

\subsection{Algorithm}

Using the assumptions detailed in \S~3.1, we simultaneously fit all
the velocity slices, using similar methods to those used by Gabel
et~al.\ (2006).  The $U$ and $\vy{N}{H}$ associated with each of the
15 velocity bins are the constrained free parameters.  All velocity
bins have the same given set of CNO abundances relative to H.  The
abundances are introduced into the fit via a linear scale factor of
the metal ionic column densities predicted by the solar metalicity
grid of models.  This is possible since all the absorbers considered
here are optically thin to the EUV ionizing radiation (i.e., the
\heii\ and \hi\ ionization edges are optically thin), thus the total
ionic column densities scale linearly with the abundances of their
parent element.  After minimizing the $\chi^2$ in each velocity bin,
we sum all the minimum values from all the velocity bins.  We repeat
this process for 11 abundances values for each element (spaced equally
in log-space), for a total of $11^3$ models.

In each velocity bin, there are five measured $N_{ion}$; the summation
of these over the 15 velocity bins gives a total of 75 model
constraints.  The number of free parameters includes two per each
velocity bin ($U$ and $\vy{N}{H}$) times 15 velocity bins plus the
three CNO abundances, all together 33 free parameters.  The best-fit
solution is then the set of velocity-dependent $U$, $\vy{N}{H}$ values and
the C, N, and O abundances that minimize $\chi^2$ from:
\begin{equation}
\chi^2 = \Sigma_i \Sigma_j [\frac{N_{obs;j,i} - N_{mod;j}(U,N_H) f_j}{\sigma_{j,i}}]^2,
\label{eq:chi2_vel_sum}
\end{equation}
where the measured column density of the $j^{th}$ ion in the $i^{th}$
velocity bin $N_{obs;j,i}$ is compared with the model column densities
from the grid $N_{mod;j}$, $f_j$ is the scale factor relative to solar
abundances, and $\sigma_{j,i}$ the measurement uncertainty.  

Our initial models sampled a cube of CNO abundance values spanning
0.1-10 times their solar values at a resolution of $\sim$60\% between
grid points. Upon finding the general location of the best set of CNO
abundances we iteratively refined the span of abundances to a final
resolution of 3\% between grid points. It is important to verify that we
found the absolute rather than a local $\chi^2$ minimun and to check
whether there are other local minima that may give a different, but
valid abundances solution.  For these purposes we ran additional
grid-models.  In particular, we ran. the initial cube model with
triple resolution and studied the behavior $\chi^2$ in that volume.
We found no other minima in that phase-space volume.

\begin{figure}[ht]
\centerline{\psfig{file=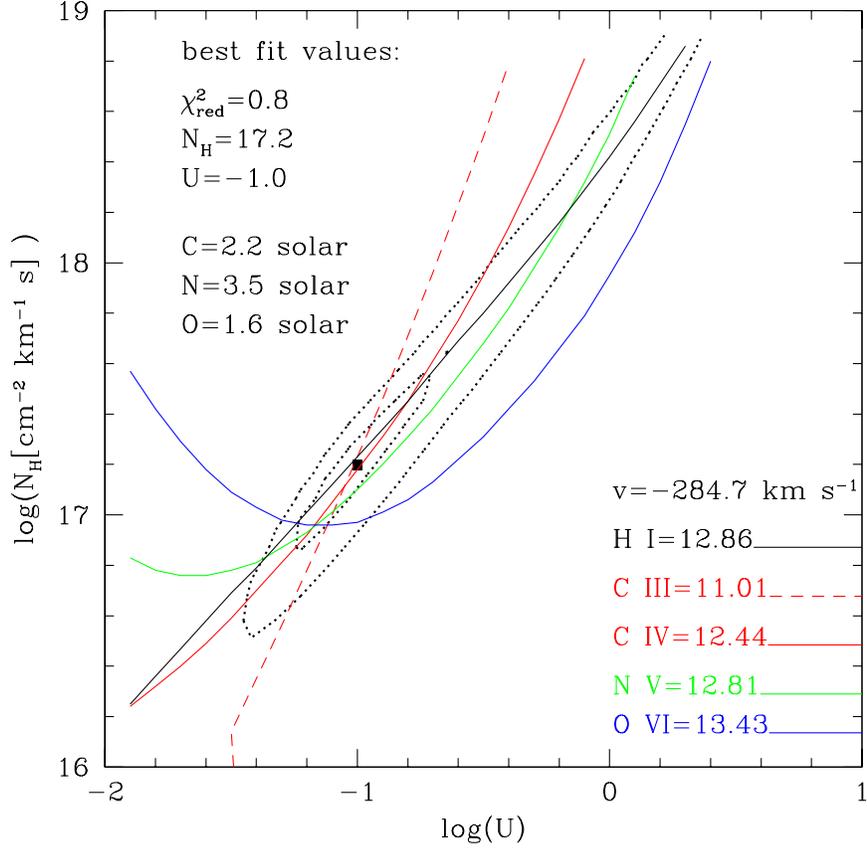,angle=0,height=12.0cm,width=12.0cm}}
\caption{ Similar to Fig.\ \ref{fig:nh_u_285_f_chi2} 
 for the best global abundances solution.
Reduced $\chi^2$ contours are plotted at 5 and 20 for an
$\vy{N}{H},U$ solution (dotted lines).  Position of the formal
solution is marked by the filled square and the best fit values are
shown at the top left corner.  The best fit reduced $\chi^2$ gives an
acceptable $\vy{N}{H},U$ solution for this abundance solution.  }
\label{fig:nh_u_285_abun2}
\end{figure}

\subsection{Abundances Results}

The best fit for all 15 velocity slices is obtained using the
following set of CNO abundances (relative to solar) carbon=2.2,
nitrogen=3.5 and oxygen=1.6.  For this model, equation
\ref{eq:chi2_vel_sum} yields $\chi^2 = 47.5$ for 41 degrees of freedom
or a reduced $\chi^2$ of $\chi_{red}^2=1.16$.  In figure
\ref{fig:nh_u_285_abun2} we show the formal $\chi^2$ solution for the
best fit values of $\vy{N}{H}$ and $U$ of the single velocity slice
shown in figure \ref{fig:nh_u_285_f_chi2} .  The reduced $\chi^2=0.8$,
formally shows that the model yields a statistically valid fit for the
data.  A similar situation occurs in most other velocity slices that
include the same abundances solution. The range of reduced $\chi^2$
in all velocity slices is 0.5--2.1, for three degrees of freedom.

We calculate the model-independent errors associated with these
abundance determinations by finding the 90\% single-parameter
confidence intervals obtained from changing only the said abundance
while keeping the others fixed (following Press et~al.\ 1989 and Taylor
1997).  The results are shown in Table 1.  
Thus, carbon and oxygen are enhanced by a factor of $\approx$ 2
relative to solar values, while nitrogen is enhanced by a factor of
3-4,  in agreement with the $Z^2$ scaling indicative of enhanced
secondary production in massive stars (Hamman et~al.\ 1999, and
references therein).

One can also find a model-dependent error-estimate for an assumed
$Z^2$ scaling, since in that model the ratio of C/H and O/H are
proportional to each other while N/H$\propto$(O/H)$^2$. For this model
the 90\% confidence level for nitrogen is only slightly lowered but
the error bars on the oxygen and carbon abundances shrink by 60\% and
50\% respectively.

\begin{deluxetable}{ll}
\tablecaption{\sc abundances of the Mrk~279 outflow relative to solar\tablenotemark{a}}
\tablewidth{0pt}
\tablehead{
\colhead{element}
&\colhead{abundance}
}
\startdata
carbon   & 2.2$\pm$0.7    \\
nitrogen & 3.5$\pm$1.1    \\
oxygen   & 1.6$^{+0.7}_{-0.8}$  \\
\enddata
\tablenotetext{a}{see \S~2.3}
\label{abundance_table2}
\end{deluxetable}


\section{AGREEMENT WITH THE X-RAY PHOTOIONIZATION RESULTS}

As noted above, we obtained simultaneous X-ray observations (360
ks. Chandra/LETGS) of the Mrk~279 AGN outflow. The full analysis of
the X-ray data is presented by Costantini et~al.\ (2006).  Kinematic
similarity between the UV and X-ray ionized absorber in this object
strongly argues that we are seeing the same outflow in both spectral
bands.  It is therefore important to compare the independently derived
X-ray ionization equilibrium findings, with those presented here for
the UV data. 

Costantini et~al.\ (2006) explored two physical scenarios for the
X-ray ionized absorber.  The first set of models postulated discrete
ionization components, and it was found that a model with two
ionization zones gives a good fit to the X-ray absorption data.  The
second set of models postulated a continuous distribution of ionization
parameter for the absorbing material.  Such models were successful in
fitting the X-ray absorption data of NGC~5548 (Steenbrugge et~al.\
2005).  Unlike the case for NGC~5548, models with a simple power-law
distribution in ionization parameter did not yield a satisfactory fit
to the Mrk~279 Chandra/LETGS data.  A bimodal distribution was needed
for an acceptable fit.  Such a model is close in concept to a two
ionization zones model.  Since the simple two ionization zones model
already gives a good fit to the data, we conclude that it provides a
more plausible physical description for the Mrk~279 X-ray absorber.
We therefore compare our UV results to the results of the two
ionization zones X-ray model.

For the two ionization components model, Costantini et~al.\ (2006)
found the following values (their table 6): \\ 
1. $\log(\xi)=0.47\pm0.07$ erg~s~cm$^{-1}$,
$N_H=1.23\pm0.23\times10^{20}$ cm$^{-2}$ \\ 
2. $\log(\xi)=2.49\pm0.07$ erg~s~cm$^{-1}$,
$N_H=3.2\pm0.8\times10^{20}$ cm$^{-2}$, \\ where the ionization
parameter $\xi$ is defined as $\xi\equiv \L/\vy{n}{H}r^2$ ($L$ is the
luminosity of the source; $\vy{n}{H}$ is the number density of
hydrogen in all forms; $r$ is the distance from the central source).
For a given SED there is a one-to-one correspondence between the
ionization parameter $\xi$ and the ionization parameter $U$, which we
are using in this paper.  For our Mrk~279 SED we find
$\log(\xi)-\log(U)=1.5$.  Therefore, for component 1 we find
$\log(U)=-1.0$.  Figure \ref{fig:nh_u_285_abun2} shows the ionization
solution for one velocity slice where $\log(U)=-1.0$. This value is
representative to all our velocity slices, as we find 90\% of the
material in this velocity range to have $-1.2<\log(U)<-0.9$.  Thus,
the ionization parameter of the low ionization component deduced from
the X-ray analysis is practically identical to the one deduced from
the UV analysis.  The high ionization X-ray component does not have
any bearing on the UV analysis since this component is not expected to
contribute measurable column density for the UV ions.  
This is easily verified by inspecting figure~\ref{fig:nh_u_285_c_nion}.
The relative fraction of the UV ions drops dramatically above $\log(U)=-1.0$.
Even the fraction of highest ionization UV species (\ovi), is lower
by more than two orders of magnitude  for  $\log(U)=1.0$ than for $\log(U)=-1.0$.

How does the total $N_H$ for the low ionization X-ray component
compare with our UV findings?  For the velocity range $-220$ \kms\ to
$-300$ \kms\ we find from the UV analysis: total
$N_H=0.8\times10^{19}$ cm$^{-2}$. In order to compare this value to
the X-ray determined $N_H$ two factors have to be taken into account.
First, the velocity interval of the UV analysis is only 80 \kms\ out
of the 340 \kms\ of the entire UV trough.  It is possible that the
rest of the trough has different amount of $N_H$ per \kms\ on average,
but in order to get a rough comparison we will simply multiply the
above UV derived $N_H$ by 340/80.  Second, the X-ray analysis was done
assuming solar abundances.  Since the UV metalicity enhancement is a
factor of 2.4 on average, we need to multiply the UV derived $N_H$ by
this additional factor.  Accounting for both factors, we find that our
rough X-ray equivalent $N_H$ is $N_H=0.82\times10^{20}$ cm$^{-2}$, which is
compatible with the derived value for the low ionization X-ray
component: $N_H=1.23\pm0.23\times10^{20}$ cm$^{-2}$.  It is remarkable that
that the fully independent X-ray and UV ionization
solutions agrees so well for the Mrk~279 absorber.

Finally, we address the issue of whether our UV derived abundances are
compatible with the X-ray analysis.  In principal, it is very
difficult to derive absolute abundances from the X-ray data since
there are no hydrogen lines in the X-ray band. Therefore, the X-ray
data are insensitive to simple metalicity scaling and it is not
surprising that Costantini et~al.\ (2006) find solar metalicity to be
adequate for their models.  Barring a handle on the absolute abundances
we look at secondary effects involving ratios of nitrogen to carbon.
Our UV findings show that the outflow's (N/C)$=1.6\times$(N/C)$_\odot$.
The best X-ray measurements are those of \cvi\ and \nvi.
From table 5 in Costantini et~al.\ (2006) we find that the modeled \cvi\
is virtually identical to the
measured \cvi.  However, the  measured  \nvi\ column density:
$\log(N[{\rm \nvi}])_{\rm obs}=16.9\pm0.6$ cm$^{-2}$ 
is higher than the modeled one
$\log(N[{\rm \nvi}])_{\rm model}=16.3\pm0.36$ cm$^{-2}$.
(although the error bars on the two values  make them formally consistent
 with each other).
Although the measurements and modeling errors are large, the X-ray
data are consistent with a relative enhancement of (N/C) compared solar, 
similar to the one deduced from the UV data.  This consistency
strengthens our UV abundances findings.

\section{DISCUSSION}

\subsection{Robustness of the Abundances Determination}

We argue that the work described here is the first reliable abundance
determinations in  AGN outflows.  To support this claim let us review
what is necessary for deriving reliable abundances in the outflows,
and how well  this project met these requirements. In \S~5.3
we describe how other previous and current efforts fall short of satisfying
these necessary conditions. 

Determining the ionization equilibrium and abundances (IEA) in an AGN
outflow depends crucially on obtaining reliable measurements of the
absorption column densities ($N_{ion}$) from the observed troughs. As
described in the Introduction and elsewhere (Arav et~al.\ 1999b;
de~Kool et~al.\ 2001; Arav et~al.\ 2001a; Gabel et~al.\ 2003; Scott et
al.\ 2004) solving for the velocity-dependent covering fraction is
crucial for obtaining reliable $N_{ion}$ from the troughs.  Column
densities inferred using other techniques (curve of growth, apparent
optical depth and Gaussian modeling) suffer from large systematic
uncertainties and many times can only be used as lower limits.  We
have developed state-of-the-art velocity-dependent methods to
determine the covering fraction of both the continuum and BELR as well
as the real optical depth (Gabel et~al.\ 2005a and this work).  To
implement these methods a suitable data set must have the following
attributes:

\noindent 1) Spectral resolution larger than 20,000 in order to allow
   for velocity-dependent analysis across the troughs. FUSE and the
   Echelle gratings on board HST give us sufficient resolution.

\noindent 2) To enable a meaningful covering factor analysis of the
   CNO doublets, the S/N of the data must be $\gtorder20$ for the
   above resolution.  The reason for such high S/N is the need
   to quantify the differences between the residual intensities 
   of two doublet troughs, which are often only 10-20\% apart 
   (see Gabel 2005a).   Very few data sets in the literature achieve
   this S/N level since it requires roughly $\gtorder15$ HST/STIS
   orbits and $\gtorder100$ ks of FUSE exposure time on the brightest
   AGN outflow targets.

\noindent 3) The data set must cover troughs from enough ions to be useful for
   IEA determinations.  At the minimum we need to cover the CNO
   doublets and three or more of the Lyman series troughs.  The latter
   is a crucial requirement since  abundances cannot be 
   determined without an accurate measurement of \hi\ column
   density.  Only a combination of FUSE and HST can cover the needed troughs.

\noindent 4) The troughs associated with different doublet components
   must be unblended.  Otherwise it is impossible to extract covering
   factors and the derived $N_{ion}$ are reduced to lower limits.
   Similarly, the outflow troughs cannot be contaminated with
   significant galactic absorption.

\noindent 5) FUSE and HST/STIS coverage must be near-simultaneous, preferably
   within the same month.  Seyfert outflow troughs are known to show
   considerable changes over time scales of more than several months
   (e.g., NGC 3783, Gabel et~al.\ 2004; Mrk 279, Scott et~al.\ 2004).  Any
   combined analysis of the full data set must rely on no changes in
   the absorption between the different epochs.

The deep and simultaneous FUSE and HST/STIS observations of Mrk~279 are
the only existing data set that satisfies all these requirements.
Therefore, our measurements yield the first sufficient set of reliable
$N_{ion}$ to allow for a physical determination of the IEA in AGN
outflows.

Our next step was to infer the physical conditions of the absorbing
gas using these $N_{ion}$ measurements.  Here we introduced an
analysis method that uses the full information imprinted on these
high-resolution data.  (In parallel, Gabel et~al.\ 2006 used similar
techniques on the VLT/UVES spectrum of QSO J2233-606.) Instead of
integrating the $N_{ion}$ across a kinematic component we treat each
resolution element separately.  

Physically, this approach is justified
since elements of the flow that are separated by more than several
\kms\ represent physically-separated and sonically-disconnected
regions.  In this way, instead of having constraints on the physical
conditions in one kinematic component, we obtain constraints on 15
separate regions.  We therefore do not have to use the implicit
assumption that the ionization equilibrium is constant across a given
component.   Working at the spectral resolution of
the spectrograph provide the maximum kinematical information
without significant scattering between adjacent spectral elements.
We also note that these spectrographs have very little instrumental 
scattered-light component (Howk \& Sembach 2000) 

We found considerable variation for $U$ and $\vy{N}{H}$ over 
the 15 velocity slices:  $-1.4<\log(U)<-0.9$ and $16.3<\vy{N}{H}<17.4$.
Thus it is evident that ``classical'' kinematic components that often covers
more than 100 \kms, indeed show significant variation in $U$ and $\vy{N}{H}$
as a function of velocity.

For one kinematic component given the set of measured $N_{ion}$, it is
always possible to find a set of CNO abundances that will produce a
perfect photoionization solution.  It is therefore impossible to
assess the physical significance of such a solution.  The situation is
different for constraining 15 separate regions. We rely upon two
plausible physical assumptions, that the individual $U$ and
$\vy{N}{H}$ can vary from one region to another, and that the
abundances across these regions is constant.  This allows us to obtain
tight constraints for the individual CNO abundances as described in
\S~3.4.  We spent a great deal of effort to derive physically
meaningful errors on our $N_{ion}$ measurements.  It is a major
achievement that the excellent statistical fit of our abundance model
(reduced $\chi^2=1.16$) was achieved without any re-adjustments of
these errors. This not only validates our methods, but also the
physical model we are using to analyze AGN outflows.

\subsection{Possible Caveats}

\subsubsection{Inhomogeneous absorber models}

As described above, we invested a large effort in developing and
utilizing analysis techniques that solved for two separate covering
factors and the optical depth as a function of velocity.  An important
question in this context is how valid is the assumption of a rigid
absorbing-material distribution (essentially a step-function) assumed
by the partial covering model?  Would an inhomogeneous distribution of
absorbing material across the emission source (de Kool, Korista \&
Arav 2002) yield a good alternative to the partial covering model?
In Arav et~al.\ (2005) we tested inhomogeneous absorber models on the
Mrk~279 data set, where the three high quality \Lya, \Lyb\ and
\Lyg\ troughs provided the strongest constraints. We concluded that
inhomogeneous absorber models that do not include a sharp edge in the
optical depth distribution across the source are not an adequate
physical model to explain the trough formation mechanism for the
outflow observed in Mrk~279. This result supports the
use of partial covering models for AGN outflows.

\subsubsection{SED uncertainties}

As stated above, the SED we use (see Fig.~\ref{fig:SED}) is arguably the 
best-determined one for an individual Seyfert study.  Nonetheless, we do not
cover the crucial 13.6--150 eV ionizing part of the SED, and therefore
there is inherent uncertainty that affects the photoionization
models. The softest plausible SED is where we connect the last
UV point with the first X-ray point (see \S~2.2) with a simple power-law. This
minimizes the assumptions regarding the shape of the unseen portion of
the SED, but with the price of assuming a spectral break immediately
following the last UV point.  We ran photoionization models using such SED
and found that the largest change was a decrease in the carbon abundance
by $\sim$10\%, with nitrogen and oxygen showing similar but smaller change.

\subsection{Comparison With Other Outflow Metalicity Determinations}

All IEA studies of Seyfert outflows prior to 2003 used $N_{ion}$
measuring techniques that are inadequate for AGN outflows troughs:
curve of growth, apparent optical depth and Gaussian modeling.  The
key importance of the covering factor was overlooked.  Therefore, the
IEA findings of these studies suffer from large unquantified
systematic errors that can only be corrected to some extent by redoing
the analysis using more accurate trough formation models.  In many
cases a reanalysis is unwarranted since one or more of the conditions
detailed in \S~5.1 is not met:  S/N of the data is
insufficient (e.g., Fields at al 2005); data do not cover troughs
from enough ions to be useful for IEA determinations  (e.g.,
Brotherton et~al.\ 2001); doublet troughs are blended (e.g., NGC 4151,
Kraemer and Crenshaw 2001); severe contamination with galactic
absorption (e.g., NGC 3783, Gabel et~al.\ 2003); FUSE and HST/STIS
coverage is not simultaneous.

A typical recent example is the study of the Mrk 1044 outflow (Fields
et~al.\ 2005), which has the following shortcomings: a) S/N of the
FUSE data is too low to allow a covering-factor analysis.  b) There is
severe contamination with galactic absorption in the FUSE band,
greatly affecting the measurements of the crucial \Lyb\ trough.  c)
FUSE and HST/STIS observations were taken 6 months apart allowing
significant changes in the absorber to occur, while it is necessary to
assume that no absorption change occurred for the sake of the
photoionization analysis.  d) There was no attempt to solve for the
velocity dependent covering factor for the STIS observed \civ\ and
\nv, which introduces large errors in their $N_{ion}$ determination.
e) The main measured FUSE lines (\ovi) are totally saturated, thus the
derived \ovi\ column density is only a lower limit.  Most importantly,
the main conclusion, that the outflow shows a metalicity of at least
5 times solar is weakened by the fact that the Lyman series troughs
are either heavily blended (\Lyb), or of too poor quality (\Lyg) for
analysis.  A possible factor of 3 increase in the \hi\ measurement is
reasonable under these conditions and will make the photoionization
model (see their Fig.~5) consistent with solar metalicity.


 \section{SUMMARY}

This paper presents a main result of a deep simultaneous UV
and X-ray spectroscopic campaign on the AGN outflow seen in Mrk
279, and is the fifth paper resulting from this campaign.
Our choice of object combined with the long observations (92
ks.\ with FUSE, 16 HST/STIS orbits and 360 ks. Chandra/LETGS), allowed
us to obtain the first reliable determination of chemical abundances
in an AGN outflow.  We find that relative to solar the abundances in
the Mrk~279 outflow are: carbon 2.2$\pm$0.7, nitrogen 3.5$\pm$1.1 and
oxygen 1.6$\pm$0.8 (\S~3)

Previous efforts to derive the abundances and ionization equilibrium
of the outflows suffered from using inadequate models for the
formation of the absorption troughs.  Therefore, their inferred ionic
column densities are unreliable, and these uncertainties are then
amplified by the photoionization models used to deduce the abundances
and ionization equilibrium.  Much of the problem can be traced to the
physical quantities that determine the shape of the observed troughs.
In the ISM and IGM, the shape of absorption troughs singularly depends
on the optical depth of the absorbing material.  This allows for a
straightforward extraction of the all-important ionic column
densities.  In contrast, AGN outflow absorption troughs are a
convolution of velocity dependent covering factor and optical
depth. These two quantities must be de-convolved if one hopes to
derive reliable column densities.  To accomplish this a data set must
have the following attributes(see \S~5.1 for full discussion):
Spectral resolution larger than 20,000 and S/N $\gtorder20$ in order
to allow for velocity-dependent analysis across the troughs; spectral
coverage of at least the CNO doublets and three Lyman series troughs,
where the troughs associated with different doublet components must be
unblended; and near-simultaneous observations of all these troughs.

We chose Mrk~279 because it is by far the best AGN outflow target in
satisfying the above requirements. Our deep campaign yielded the
necessary high S/N data to do the analysis properly, and avoid many
compromising assumptions that plagued previous studies.  As a result
we obtained the best determination of ionization equilibrium to date
and the first reliable measurement of abundances in these
environments.  It is also the first time that the analysis of the UV
data is in good agreement with the X-ray analysis.  That is, the
physical properties of the low ionization component seen in the X-ray
match very well to those we find for the UV gas (see \S~4).

In order to extract the physics from the data we abandoned the old
notion of ``absorption trough components.'' This notion is somewhat
suitable for ISM and IGM clouds but is not adequate for studying
dynamical absorption structures.  Instead, we developed velocity-dependent
analysis technique, which proved crucial in deriving the results of this paper.
We believe that these techniques are the only reliable way to determine
the physical properties of AGN outflows.  This applies to both 
future UV observations, as well as ground based Echelle spectroscopy of
the outflows.

This study demonstrates that the quality of science that can be achieved
by a well designed campaign is far higher than the combined results of
many small projects in this field.  Providing a strong argument for investing 
the large resources needed for such studies.



\section*{ACKNOWLEDGMENTS}

This work is based on observations obtained with {\em HST} and {\em
FUSE}, both built and operated by NASA.  The FUSE mission is operated
by the Johns Hopkins University under NASA contract NAS5-32985.
Support for this work was provided by NASA through grants number {\em
HST}-AR-9536, {\em HST}-GO-9688, from the Space Telescope Science
Institute, which is operated by the Association of Universities for
Research in Astronomy, Inc., under NASA contract NAS5-26555, and
through {\em Chandra} grant 04700532 and by NASA LTSA grant 2001-029.
SRON is supported
financially by NWO, the Netherlands Organization for Scientific
Research.
We also thank the anonymous referee for useful comments and suggestions.

\section*{Appendix: Profile Fitting and Ionic Column Density Measurements}

  In Gabel et~al.\ (2005a; hereafter G05), we performed a detailed
global fitting of the intrinsic absorption profiles in the 2003 {\it
FUSE} and STIS spectra of Mrk 279 to derive the velocity-dependent
ionic column densities and line-of-sight covering factors.  This
fitting method allowed for treatment of the individual covering
factors of the physically distinct background emission sources, i.e.,
the continuum source ($C_c$) and emission-line region ($C_l$).  While
the solution from this analysis matched the observed profiles well
overall (see Figure 6 in G05), there were some
systematic discrepancies indicating the need for a refinement in the
fitting assumptions. Given the crucial importance of having accurate
ionic column densities with realistic uncertainties for
photoionization modeling of the outflow, we re-address those
measurements below.

  Two key findings from our global fitting analysis provide the
basis for our refined measurements presented here:

$\bullet$ First, the continuum source was shown to be fully occulted
by the outflow, with non-unity effective covering factors due entirely
to partial coverage of the emission-line region.  We consider this
result to be robust: $C_c \approx$1 was found systematically over the
numerous, independently fit velocity bins associated with the highly
resolved absorption profiles, and it was found consistently in
separate fits to the CNO doublets and the Lyman series lines.  It is
also geometrically consistent with our understanding of the different
size-scales of the emission regions, with the UV continuum source much
smaller than the BLR.

$\bullet$  Second, fits to the \civ\ and \nv\ doublets deviated 
slightly, but systematically over much of the profiles,
with the absorption strength underestimated (overestimated) in the blue (red) 
members of each doublet.  This indicates the effective covering factors for 
these ions are lower than the global fit solution.
In deriving that solution, we assumed all ions share
the same covering factors -- this was necessary to sufficiently constrain
the model so that $C_c$ and $C_l$ could be treated separately.
However, with the result that $C_c=$1, we can now eliminate that simplifying 
assumption and solve the ionic column densities independently for each doublet 
pair.

   To determine column densities we used $\chi^2$ fitting analysis, 
comparing the observed normalized fluxes with the model absorption equation:
\begin{equation}
I = 1 + (R_l C_l + R_c C_c) (e^{-\tau} - 1),
\label{eq:covering}
\end{equation}
where $R_l$ ($R_c$) are the fractional flux contributions of the line
(continuum) emission sources underlying the absorption and $\tau$ is
the line optical depth.  Equation \ref{eq:covering} comes from the more general
expression for an arbitrary number of emission components given in
equation 4 of G05 (see also Ganguly et~al.\ 1999).  For each doublet
pair, $C_l$ and $\tau$ were derived in each velocity bin using
equation \ref{eq:covering} with $C_c=$1 by minimizing the $\Delta\chi^2$ function summed
over the two doublet lines; the optical depths of the two lines are
related by their intrinsic 2:1 ratio.  The emission-line covering
factor was constrained to the physically meaningful range of 0$\leq
C_l \leq$1.  Optical depths were converted to ionic column densities
using the expression in Savage \& Sembach (1991).  For 1$\sigma$
uncertainties in these parameters, we adopt the maximum offsets from
the best-fit values which give $\chi^2 =$2 (i.e., equal to the number
of lines being fit).  For cases of large optical depth (typically
$\tau \gtrsim$4), no upper limit can be determined because of heavy
saturation.  In these cases, there are only lower limits on the ionic
column densities.  We measured the \hi\ column density from the
Lyman series lines in a similar way.  In this case, we fit the
combined Ly$\alpha$, Ly$\beta$, and Ly$\gamma$ lines since they all
exhibit strong absorption and have the highest signal-to-noise ratio.
Since there are three lines in the Lyman fitting, 1$\sigma$
uncertainties in the fitted parameters are values giving $\Delta\chi^2 =$3.

\begin{figure}[h]
\epsscale{0.5}
\plotone{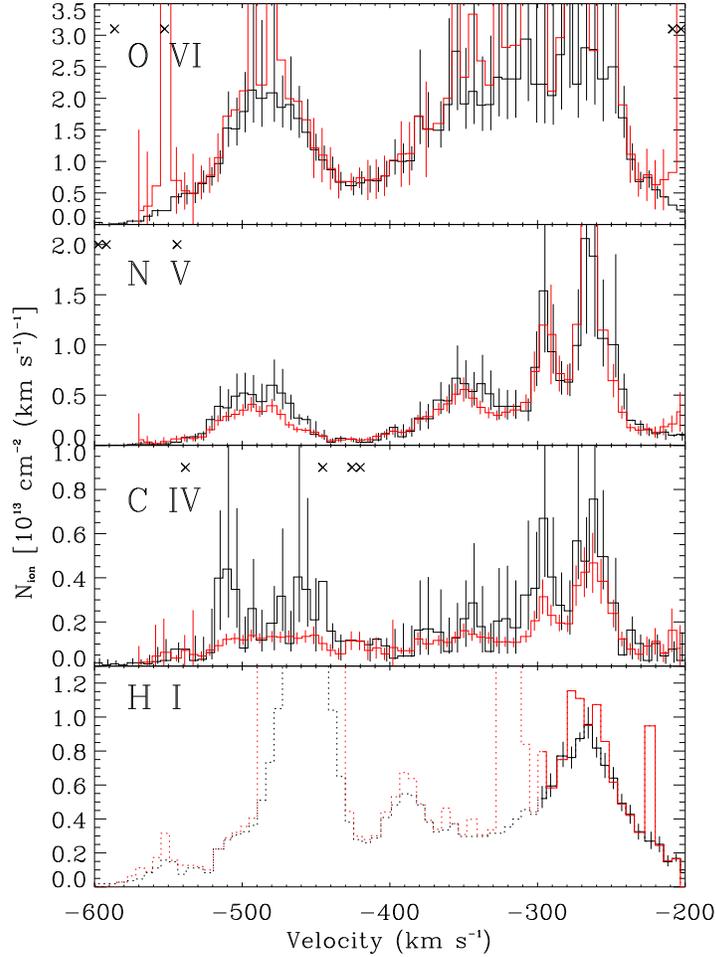}
\vspace*{0.5 in}
\caption{Ionic column density profiles derived for the UV outflow in Mrk 279.
Results from $\chi^2$ minimization fitting of the absorption profiles,
with $C_c=$1 as described in the text, and associated 1 $\sigma$ 
uncertainties are shown with black histograms and error bars. 
For comparison, the global-fit solutions from G05 are also plotted
(red).  Velocity bins with no valid solution (minimum $\Delta\chi^2 \geq$ 2)
are denoted with $\times$.  The region in the \ion{H}{1} profile with blended
absorption from different physical components and thus no solution is
plotted with dotted histograms (see G05).}
\label{fig:N_ion_errors}
\end{figure}

\begin{figure}
\epsscale{0.7}
\plotone{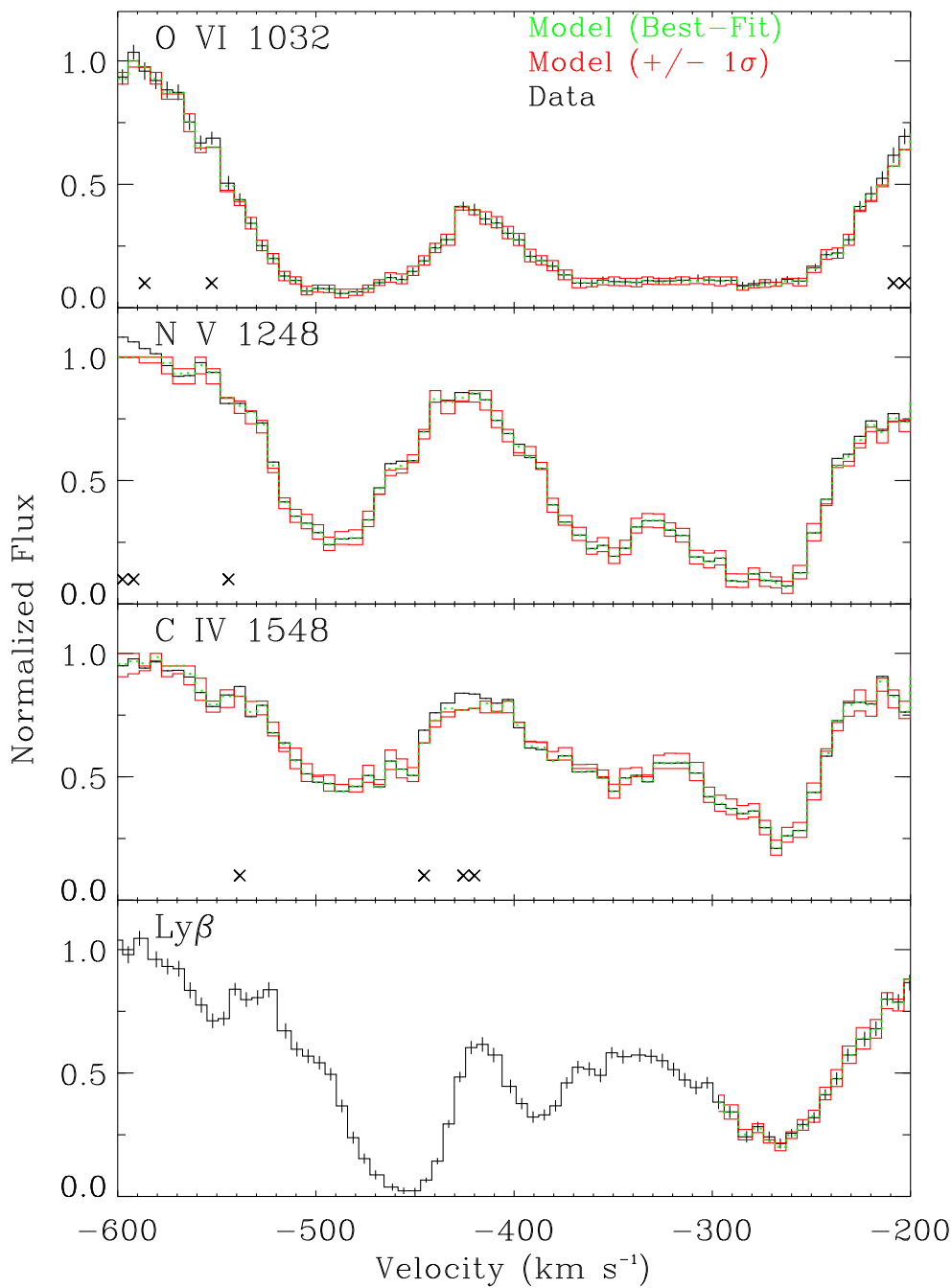}
\vspace*{0.5 in}
\caption{Absorption profile fits and uncertainties compared with the
observed spectra. Model profiles corresponding to the best-fits to
$\tau$ and $C_l$, with $C_c=$1, are shown in green.  Profiles derived
using $+$/$-$ 1$\sigma$ uncertainties in the fitted parameters are
plotted (red) for comparison with the 1$\sigma$ uncertainties in the
fluxes (black). Velocity bins with no valid solution are marked with
$\times$ (see G05).}
\label{fig:trough_profile_model_errors}
\end{figure}

   The column density profiles derived from our $\chi^2$ fitting are
shown in figure \ref{fig:N_ion_errors}, together with the global fit
solutions from G05 for comparison.  This shows some important
differences that would affect the results of photoionization models.
Over much of the profiles, the \ion{N}{5} and, especially, \ion{C}{4}
column densities are larger than in the global fit, due to the lower
effective covering factors associated with these ions.  Also, the
estimated uncertainties associated with the $\chi^2$ analysis, which
we consider to be more realistic as described above, are generally
larger.  Figure \ref{fig:trough_profile_model_errors} shows the model profiles compared with observed
fluxes.  Profiles generated using $+$/$-$ 1$\sigma$ uncertainties in
the parameters (red) are seen to match the flux measurement
uncertainties (black error bars) well overall.


\clearpage


\end{document}